\documentclass[10pt,a4paper,twoside]{article}
\usepackage{epsfig}
\usepackage{baltlat5}
\usepackage{wrapfig}
\begin{document}
\ \
\vspace{-0.5mm}

\setcounter{page}{1}
\vspace{-2mm}

\titlehead{Baltic Astronomy, vol.\ts 17, 223-234, 2008.}

\titleb{IRC$-$10443: A MULTI-PERIODIC SRa VARIABLE AND THE NATURE OF LONG SECONDARY PERIODS IN AGB STARS}

\begin{authorl}
\authorb{U.~Munari}{1}
\authorb{A.~Siviero}{1}
\authorb{P.~Ochner}{2}
\authorb{S.~Dallaporta}{2}
\authorb{C.~Simoncelli}{2}
\end{authorl}

\begin{addressl}

\addressb{1}{INAF Osservatorio Astronomico di Padova, via dell'Osservatorio
8, 36012 Asiago (VI), Italy}

\addressb{4}{ANS Collaboration, c/o Osservatorio Astronomico, via
dell'Osservatorio 8, 36012 Asiago (VI), Italy}

\end{addressl}

\submitb{Received 2008 August 30; revised September 3; accepted September 8}

\begin{summary}

We obtained $B$$V$$I_{\rm C}$ photometry of IRC$-$10443 in 85 different nights
distributed over two years, and in addition low resolution absolute
spectro- photometry and high resolution Echelle spectroscopy. Our data show
that IRC $-$10443, which was never studied before in any detail, is a SRa
variable, characterized by $\Delta B$=1.27, $\Delta V$=1.14 and $\Delta
I$=0.70~mag amplitudes and mean values $<$$B$$>$=13.75, $<$$V$$>$=11.33 and
$<$$I_{\rm C}$$>$=6.18~mag. Two strong periodicities are simultaneously
present: a principal one of 85.5 ($\pm$0.2) days, and a secondary one of 620
($\pm$15) days, both sinusoidal in shape, and with semi-amplitudes $\Delta
V$=0.41 and 0.20~mag, respectively. IRC$-$10443 turns out to be a M7III
star, with a mean heliocentric radial velocity $-$28~km/s and reddened by
$E_{B-V}$=0.87, a third of which of circumstellar origin. The same 0.5~kpc
distance is derived from application of the appropriate period-luminosity
relations to both the principal and the secondary periods. The long
secondary period causes a sinusoidal variation in color of 0.13~mag
semi-amplitude in $V-I_{\rm C}$, with IRC$-$10443 being bluest at maximum
and reddest at minimum, and with associated changes in effective temperature
and radius of 85~K and 6\%, respectively. This behavior of colors argues 
in favor of a pulsation nature for the still mysterious long secondary periods
in AGB stars.

\end{summary}

\begin{keywords}
stars: pulsations -- stars: variables -- stars: AGB
\end{keywords}

\sectionb{1}{INTRODUCTION}

IRC$-$10443 (= RAFGL 2209 = NSV 11129 = BD$-$12$^\circ$5123) is a bright
($K$=1.8 mag) infrared source discovered during the Two Micron Sky Survey
(Neugebauer and Leighton 1969), that lies in the general direction of the 
Scutum Star Cloud. IRC$-$10443 was detected by the AFGL survey
(Price and Murdock 1983) at 4.2~$\mu$m, and by IRAS satellite at 12 and 25
$\mu$m. Its 2MASS magnitudes and colors are $K_s$=1.92, $J-H$=1.35,
$J-K$=1.80. Its spectral type is reported to be M6 by Neckel (1958) and
Hansen and Blanco (1975), and M6.5 by Nassau et al. (1956). 
IRC$-$10443 is present in the NSV catalog of suspected variables because $I$-band
observations, obtained at five different epochs (from 21-08-1963 to
28-06-1965) which are reported in the IRC catalog, seem to trace a
variation from magnitude 6.4 to 6.9 (however, the uncertainty of the single
measurement is similar to the dispersion of the five IRC measurements around
their mean).
Not much more is known about IRC$-$10443 and its nature. In this paper we
report on our $B$$V$$I$ photometric monitoring (85 nights distributed over
two years) and optical spectroscopic observations (low and high resolution)
of this object, and how they constrain its basic properties.

    \begin{table}
      \centering
      \includegraphics[height=18.5cm]{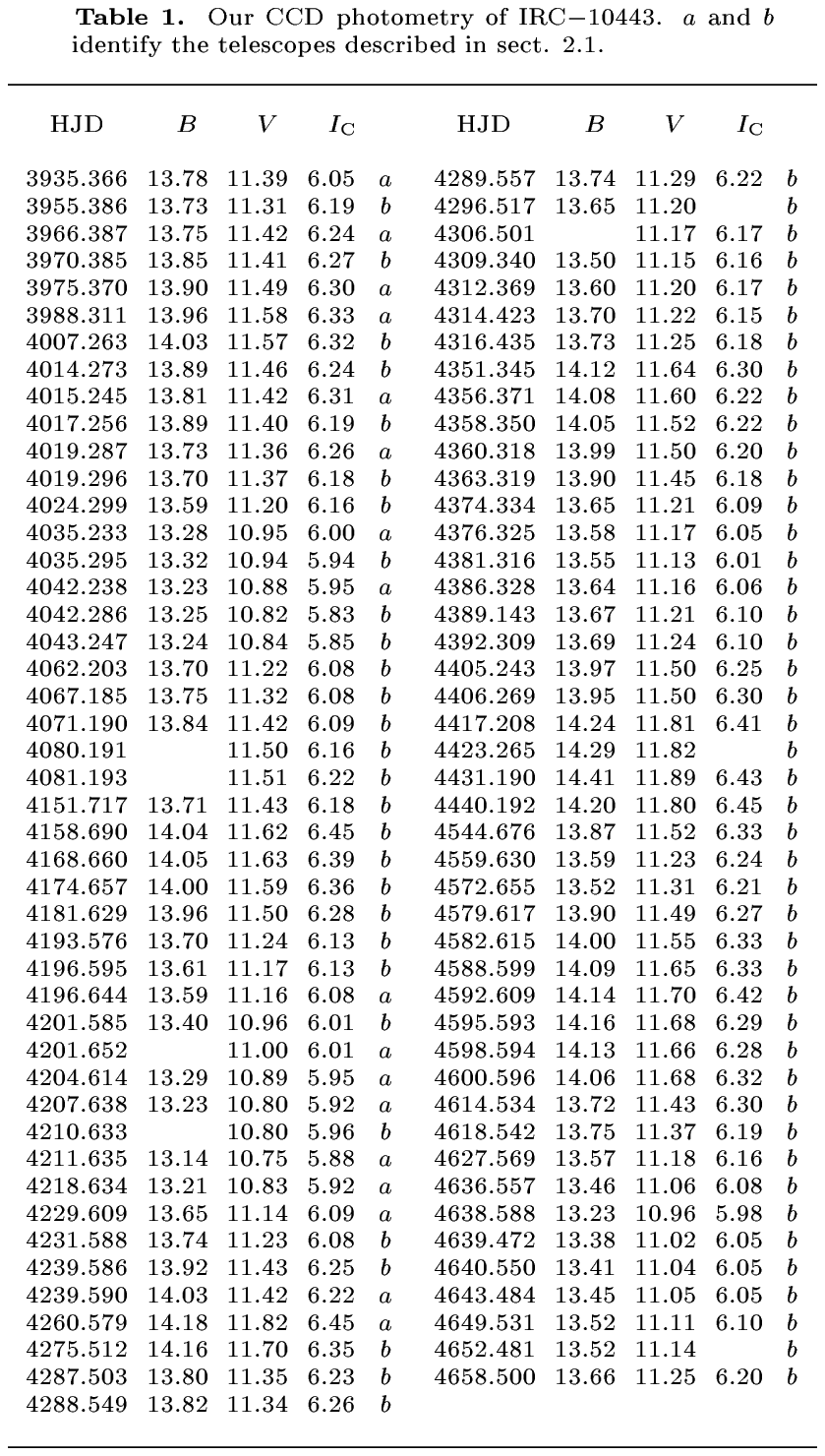}
    \end{table}

\sectionb{2}{OBSERVATIONS}
\vskip -3mm
\subsectionb{2.1}{Photometry}
$B$$V$$I_{\rm C}$ CCD photometry of IRC$-$10443 was independently obtained with two
separate telescopes: ($a$) the 0.30-m Meade RCX-400 f/8 Schmidt-Cassegrain telescope
owned by Associazione Astrofili Valle di Cembra (Trento, Italy). The CCD was
a SBIG ST-9, 512$\times$512 array, 20 $\mu$m pixels
$\equiv$1.72$^{\prime\prime}$/pix, with a field of view of
13$^\prime$$\times$13$^\prime$. The $B$ filter was from Omega and the
$V$$I_{\rm C}$ filters from Custom Scientific; and ($b$) the
0.50-m f/8 Ritchey-Cretien telescope operated on top of Mt. Zugna by Museo
Civico di Rovereto (Trento, Italy) and equipped with Optec $B$$V$$I_{\rm C}$ 
filters. The CCD was an Apogee Alta U42 2048$\times$2048
array, 13.5 $\mu$m pixels $\equiv$ 0.70$^{\prime\prime}$/pix, with a field
of view of 24$^\prime$$\times$24$^\prime$.

The comparison star for $B$ and $V$ bands was TYC~5699-6341-1, for which we
adopted $B$=11.016 and $V$=10.334 in the standard Johnson $U$$B$$V$ system.
They were obtained from Tycho $B_T$,$V_T$ data following Bessell (2000)
transformations. The comparison star for $I_{\rm C}$ band was TYC
5699-6348-1 for which we adopted $I_{\rm C}$=6.39, $V-I_{\rm C}$=0.55 from
the Hipparcos catalog. We had no alternatives for the comparison stars. In
fact, these two are the only stars within the CCD field of view of
IRC$-$10443 that ($i$) have reference magnitudes available in literature,
($ii$) are bright enough to be well exposed on the single CCD image without
risking to saturate IRC$-$10443, and ($iii$) are photometrically stable to
better than 0.02~mag. The last point was verified by noting that on all
frames in any band we obtained, the relative magnitude of the two comparison
stars was stable at this level.
 
All photometric measurements were corrected for instrumental color equations
derived nightly by observations of Landolt (1992) standard fields. The good
consistency of the data obtained independently with two different
instruments reinforce our confidence in the accuracy of the results, in
spite of the very red colors of IRC$-$10443, that are not reached by
typical Landolt standard stars. Our photometry is presented in Table~1. It
covers the period from 16-07-2006 to 11-07-2008, with observations collected
in 85 different nights. The Poissonian component of the total error budget
is less than 0.01~mag for all the data. The r.m.s. of the Landolt
standard stars around the color equations they contributed to calibrate was
on the average 0.019~mag for $B$, 0.022 for $V$ and 0.031 for $I_{\rm C}$ bands.

\subsectionb{2.2}{Spectroscopy}
A low resolution, absolutely fluxed spectrum of IRC$-$10443 was obtained on
June 24.97, 2008 UT with the B\&C spectrograph of INAF Astronomical
Observatory of Padova attached to the 1.22m telescope operated in Asiago by
the Department of Astronomy of the University of Padova. The slit, aligned
with the parallactic angle, projected onto 2 arcsec on the sky, and the
total exposure time was 1860~sec. The detector was an ANDOR iDus 440A CCD
camera, equipped with a EEV 42-10BU back illuminated chip, 2048$\times$512
pixels of 13.5~$\mu$m size. A 300 ln/mm grating blazed at 5000~\AA\ provided
a dispersion of 2.26~\AA/pix and a covered range extending from 3250 to
7890~\AA.

High resolution spectra of IRC$-$10443 were obtained on June 10.04 and July
22.95 2008 UT with the Echelle spectrograph mounted on the 1.82m telescope
operated in Asiago by INAF Astronomical Observatory of Padova. The detector
was a EEV~CCD47-10 CCD, 1024$\times$1024 array, 13 $\mu$m pixel, covering
the interval 3600$-$7300~\AA\ in 31 orders. A slit width of 200~$\mu$m
provided a resolving power $R_P$=26\,000.

\begin{figure}
\centerline{\psfig{figure=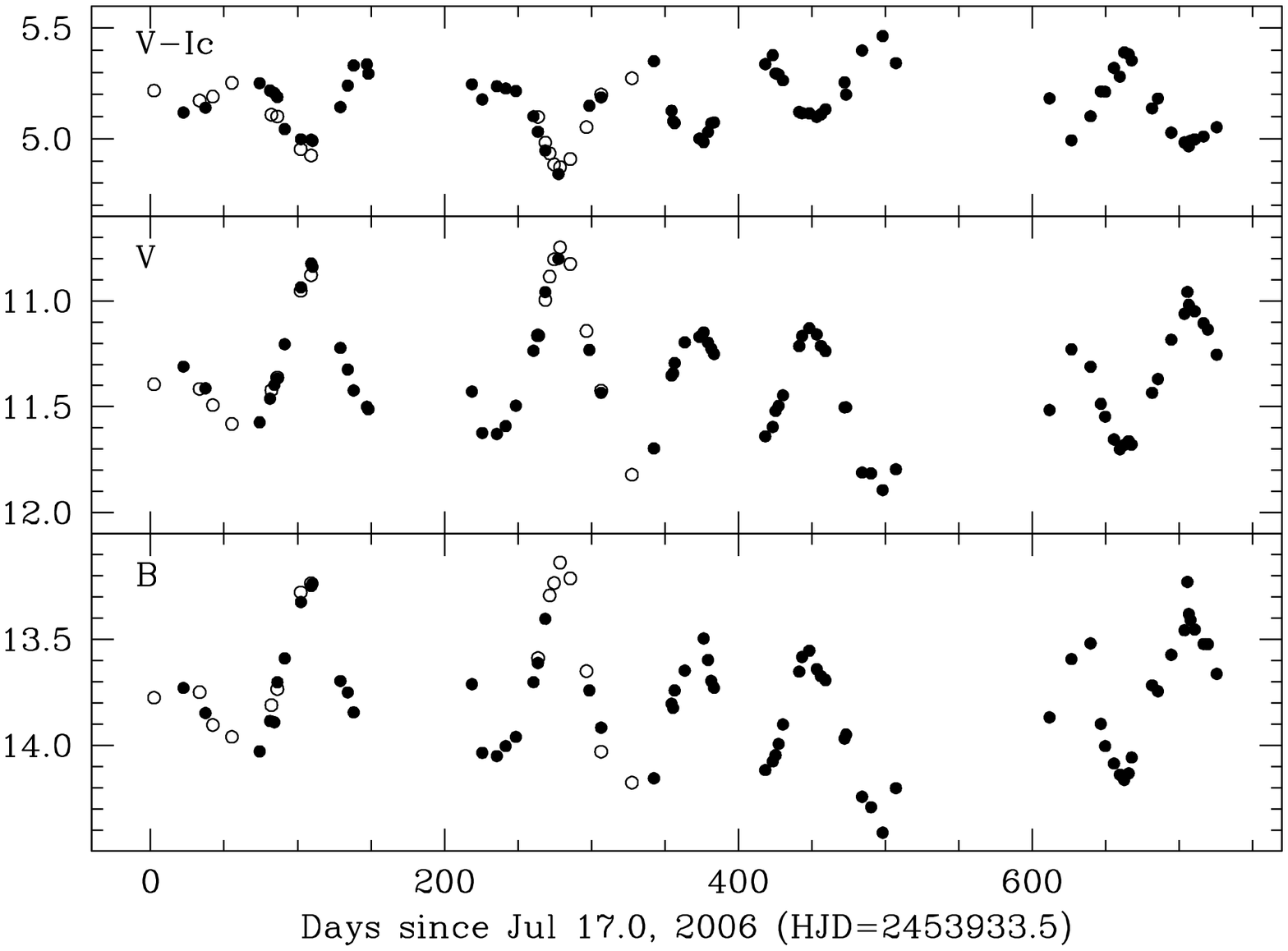,width=12.5truecm,angle=0,clip=}}
\captionc{1}{Light- and color-curves of of IRC~10443. Open circles: telescope
     $a$; filled dots: telescope $b$ (see sect. 2.1).}
\end{figure}

\sectionb{3}{RESULTS}
\vskip -3mm
\subsectionb{3.1}{Photometric variability}
The light-curve presented in Figure~1 clearly shows that IRC$-$10443 is
indeed variable. The recorded variability amounts to $\Delta B$=1.27 (from
14.41 to 13.14), $\Delta V$=1.14 (from 11.89 to 10.75) and $\Delta I_{\rm C}$=0.70
(from 6.45 to 5.75), around the mean values $<$$B$$>$=13.75, $<$$V$$>$=11.33 
and $<$$I_{\rm C}$$>$=6.18~mag.

The variability is obviously periodic (see next section) and Figure~1 shows
that the stars gets hotter (bluest $V-I_{\rm C}$) at $V$ maxima, and cooler (reddest
$V-I_{\rm C}$) at $V$ minima. This behavior of the color is a distinctive features
of stellar pulsation.

\begin{wrapfigure}[21]{l}[0pt]{65mm}
\vbox{
\centerline{\psfig{figure=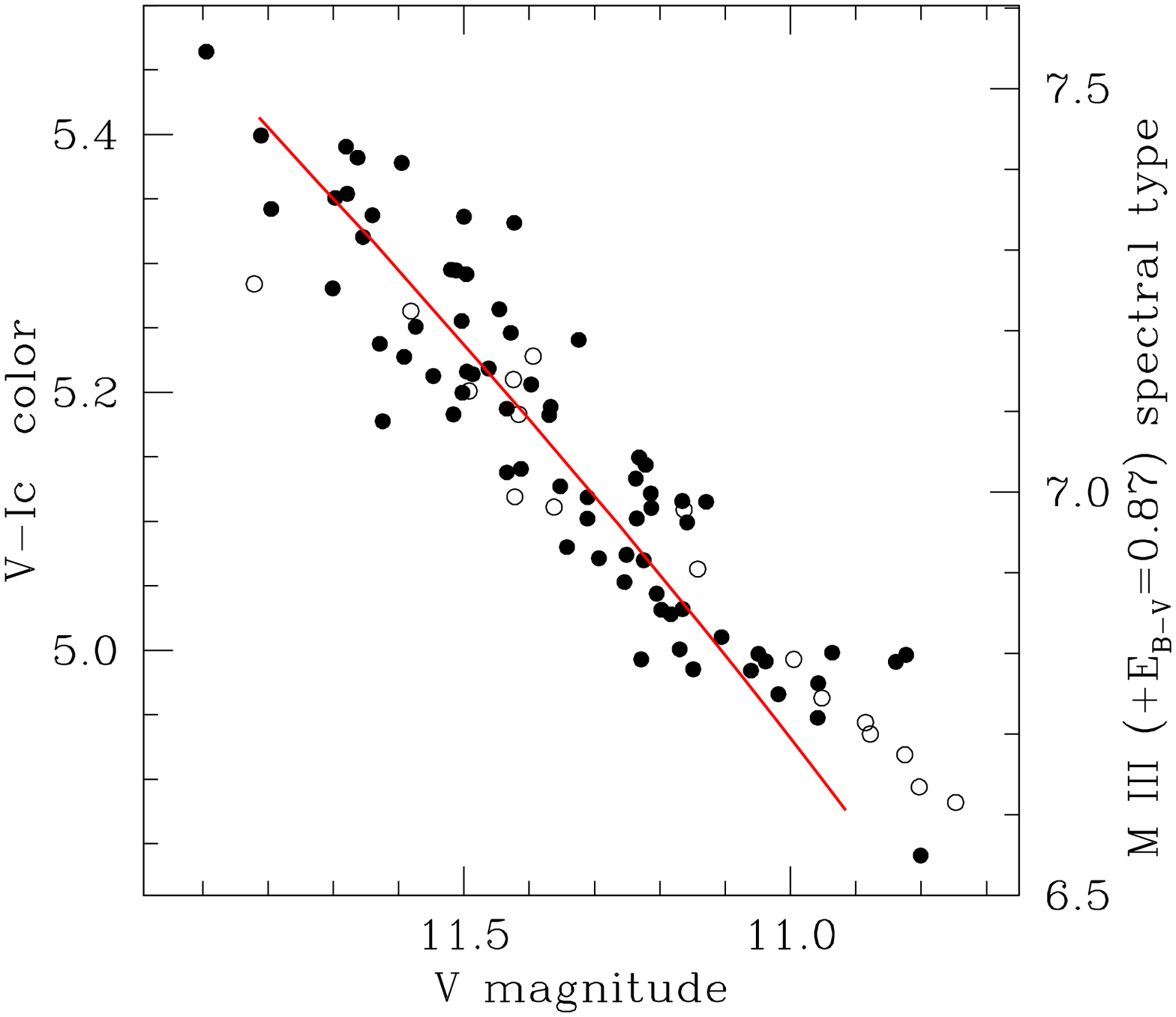,width=65truemm,angle=0,clip=}}
\vspace{-1mm}
\captionc{2}{Brightness-color diagram for the observations in Figure~1.
      The line is the path followed by a black-body, reddened by $E_{B-V}$=0.87,
      varying in radius and temperature at constant luminosity.}
}
\end{wrapfigure}

In fact, Figure~2 plots the $V-I_{\rm C}$ color against the $V$ magnitude,
showing the clear correlation between them. If we take a black-body,
redden it by $E_{B-V}$=0.87, scale its flux so to match the average $V$
band brightness of IRC$-$10443, and let it varies at constant luminosity, we
obtain the line in Figure~2, which is a reasonably good fit to the observed
points. This indicates that the variability displayed by IRC$-$10443 occurs
at constant luminosity, in the form of expansion + cooling and contraction +
warming, as expected in radial pulsations. The corresponding variation in
spectral type goes from M6.6~III to M7.5~III as indicated on the right
ordinate axis of Figure~2. These corresponding spectral types have been
obtained by integrating the $V$ and $I_{\rm C}$ bands on the Fluks et al.
(1994) spectra of M~III giants reddened by $E_{B-V}$=0.87 (the amount 
affecting IRC$-$10443, see sect.~3.4)

IRC$-$10443 appears as a bona fide SRa variable. SRa variables are
semi-regular late-type (M, C, S or Me, Ce, Se) giants displaying persistent
periodicity and usually small ($\Delta V$$<$2.5 mag) light amplitudes.
Amplitudes and light-curve shapes generally vary and periods are in the
range 35-1200 days. Many SRa differ from Miras only by showing smaller
light amplitudes (Whitelock 1996).

\subsectionb{3.2}{Multi periodicities}
Two main periodicities are at the same time present in IRC$-$10443: a principal and larger
amplitude variation modulated by a 85.5 day period, and a secondary and smaller one of 620
days. The following expression corresponds to the curve fitting the $V$-band
data in Figure 3 (where $t$ is in HJD $-$  2450000):
\begin{displaymath}
V(t) = 11.38(\pm0.02) + 0.41(\pm0.02)\sin \frac{t - 4042.0(\pm0.3)}{85.5(\pm0.2)} +
\end{displaymath} 
\begin{equation}
\phantom{V(t) = 11.38(\pm0.02)} + 0.22(\pm0.02)\sin \frac{t - 4141(\pm10)}{620(\pm15)}
\end{equation}
The combination of these two plain sinusoids provides a reasonably close
fitting to the observed light-curve. Nevertheless, the residuals are larger
than the observational errors, and an additional weaker component (either
periodic or irregular) is probably present. Our present data are
insufficient to characterize such an additional component, and to resolve it
a much longer photometric monitoring is required, which we plan to pursue.

\subsectionb{3.3}{Spectral classification and radial velocity}
The low resolution spectrum we obtained of IRC~$-$10443 was compared with
the digital spectral atlas of Fluks et al. (1994), that includes all
spectral types from M0 to M10 with spectra covering the whole optical range.
They are of high flux accuracy and of a resolution similar to ours.
Literature data suggest a M6/M6.5 spectral type for IRC$-$10443, but our spectrum
is quite poorly fitted by the M6III reference spectrum from Fluks et al.
(1994) library, while the match is perfect with a M7III spectrum, as shown
in Figure~4. In view of the pulsation activity present in IRC$-$10443, the
difference between our and other spectral classifications present in
literature can be accounted for by the changes in surface temperature that
characterize the pulsation activity (see right hand-side ordinates of Figure~2).

\begin{figure}
\centerline{\psfig{figure=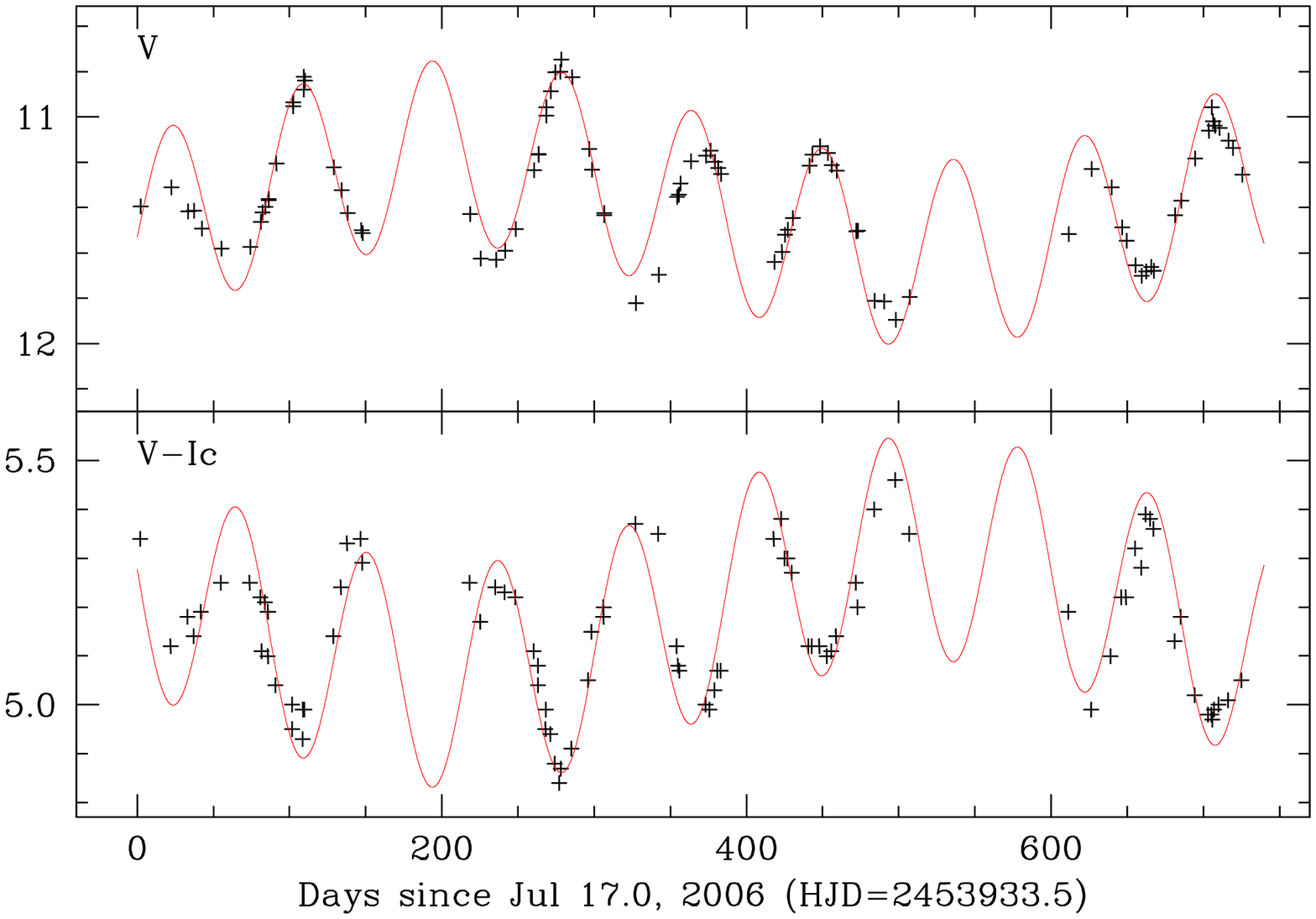,width=125truemm,angle=0,clip=}}
\captionc{3}{Fitting of the $V$ and $V-I_{\rm C}$ light-curves of IRC$-$10443 with two
      sinusoids of 85.5 and 620 days periods (see sects. 3.2, 3.6 and Eq.~1).}
\end{figure}

Figure~5 displays a portion centered on H$\alpha$ of our high resolution
Echelle spectrum of IRC$-$10443 for July 22.95, and by comparison those of
bracketing spectral types from the atlas of Bagnulo et al. (2003), degraded
to the resolution of our Echelle spectrum via a Gaussian filter. The
spectral progression in Figure~5 confirms the M7III classification for
IRC$-$10443.

Mira variables displays emission lines, mainly the higher lines in the
Balmer series, peaking in intensity at maximum brightness for both O- and
C-rich varieties (e.g. Panchuk 1978, Yamashita et al. 1977, Mikulasek and
Graf 2005), with large excursion in intensity along the pulsation cycle.
When the first Echelle spectrum was exposed on June 10, IRC$-$10443 was on
the rise and close to maximum brightness (pulsation phase 0.86), while for
the July 22 spectrum it was declining and close to minimum brightness
(pulsation phase 0.37). Both spectra do not show emission in the H$\alpha$
line, in agreement with the fact that the presence of emission lines is far
less frequent in SR than in Mira variables.

The radial velocity of the M7III star is $-$24.1($\pm$0.8) and
$-$31.4($\pm$0.7)~km/s on the June 10 and July 22 spectra, respectively. The
difference is well accounted for by the pulsation activity. The two
observations are separated in time by exactly half of the main 85.5~day
pulsation period, and their mean value $-$28~km/s can be taken as
representative of the systemic velocity of IRC$-$10443. At its galactic
coordinates ($l$=20$^\circ$, $b$=$-$3$^\circ$) and distance (0.5~kpc, see below), the
radial velocity expected from galactic disk rotation is +7~km/s (cf also
Brand and Blitz 1993).  The 35~km/s difference with the observed radial
velocity, suggests that IRC$-$10443 does not belong to the young disk
galactic population onto which it is seen projected and instead it is
related to an older population. This is confirmed by the high tangential 
velocity, 86 km/s, derived from the proper motion listed in the NOMAD catalog 
(Zacharias et al. 2004) and the distance estimated in sect.~3.5 below.

\subsectionb{3.4}{Reddening}
The fit with the  Flucks et al. (1994) M7III reference spectrum presented in
Figure~4 constrains the reddening affecting IRC$-$10443. The best match
is obtained with $E_{B-V}$=0.87$\pm$0.02 for a standard $R_V$=3.1 reddening
law.

The intrinsic $B-V$ color of M giants does not depend from the spectral type
and hence the effective temperature, as illustrated by Johnson (1966), Lee
(1970) and Fitzgerald (1970). Their tabular compilations provide
$<$$(B-V)_\circ$$>$=+1.544 as the mean intrinsic color of M5 to M8 class III
giants. The mean $B-V$ color of our observations is $<(B-V)>$=+2.418,
which corresponds to a reddening $E_{B-V}$=0.87$\pm$0.04 affecting IRC~10443.

These two independent methods converge to the same amount of reddening
affecting IRC~10443, $E_{B-V}$=0.87$\pm$0.03, which is adopted in this
paper.  A fraction of this total reddening is probably of circumstellar
origin, as supported by the detection by Kwok et al. (1997) of emission from
circumstellar dust in IRAS low resolution spectra. IRC$-$10443 lies in the
general direction of the Scutum Star Cloud (SSC, centered at $l$=27$^\circ$,
$b$=$-$3$^\circ$ galactic coordinates). SSC is one of the regions of the
Milky Way with the highest stellar density, caused by unusually low
extinction over its area. Reichen et al. (1990) presented the results of a
detailed investigation of the extinction over the SSC based on ground-based
and balloon-born UV survey data. Toward the direction to IRC$-$10443 they
found that the interstellar reddening linearly increases with distance until
$E_{B-V}$$\sim$0.55 is reached at 0.5~kpc. Longward, the further rise of the
reddening with distance is very slow, reaching $E_{B-V}$$\sim$0.65 at 4~kpc.
Only for distances $d$$>$6~kpc the reddening increases to $E_{B-V}$$\sim$1
(Madsen and Reynolds 2005).

Following the results of Reichen et al. (1990), we therefore conclude that
$\sim$1/3 of the total $E_{B-V}$=0.87 reddening affecting IRC$-$10443, is of 
probable circumstellar origin.

\subsectionb{3.5}{Distance}
In a seminal paper, Wood et al. (1999) used MACHO observations of late-type
giant variables in LMC to produce a period-luminosity diagram for them,
where five separate period-luminosity sequences were identified. Comparing
with the model prediction of Wood and Sebo (1996), three of these sequences
were found to coincide with the fundamental and first overtones pulsation
modes, while the other two seemed to trace the variability induced by
ellipsoidal distortion of RGB and AGB giants harbored in binary systems.

\begin{figure}
\centerline{\psfig{figure=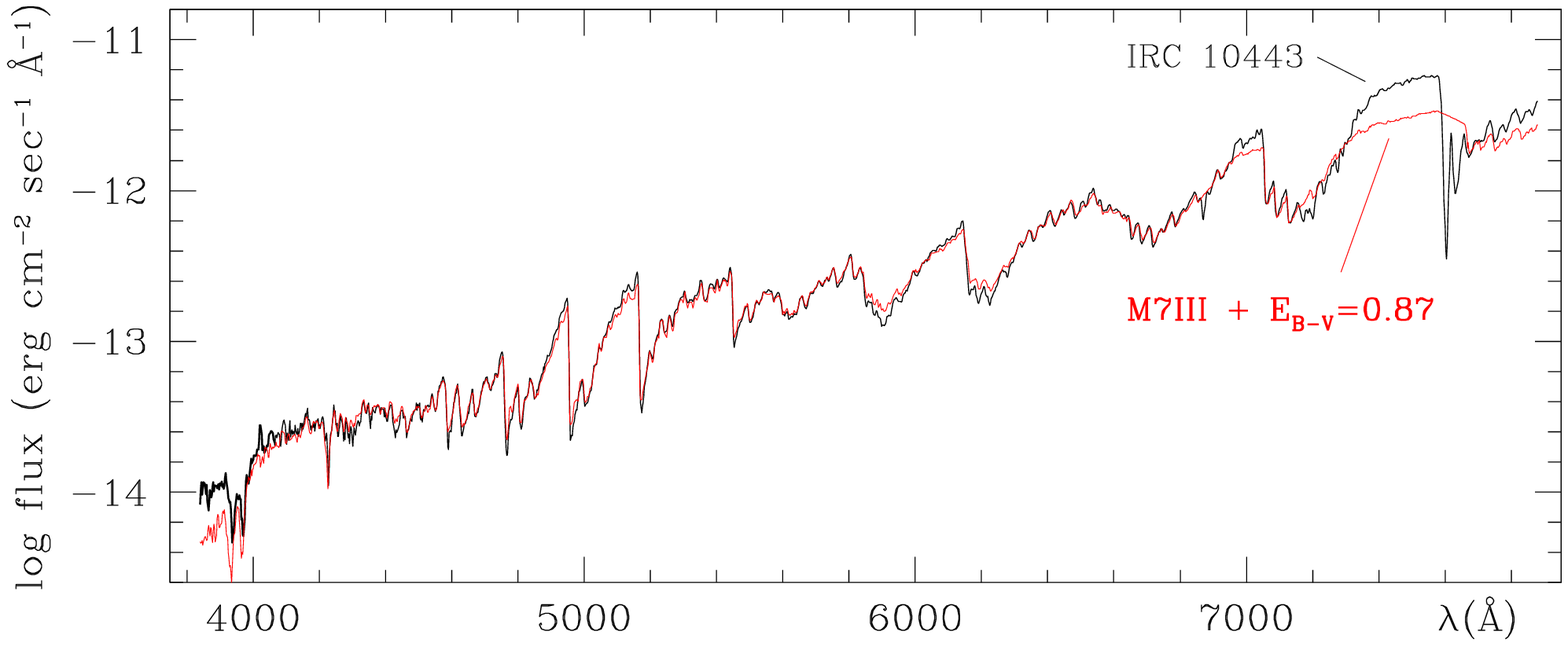,width=125truemm,height=60truemm,angle=0,clip=}}
\captionc{4}{Absolute spectro-photometry of IRC$-$10443 for 24.97 June 2008 UT
     (thick line) with superimposed the reference spectrum of a M7III star from
     the atlas of Fluks et al. (1994) reddened by $E_{B-V}$=0.87 (thin line).} 
\end{figure}

Since then, the availability of huge sets of data from large micro-lensing
surveys (MACHO, OGLE, EROS, MOA) contributed to rapidly refine the picture,
with now up to 14 different period-luminosity sequences identified (e.g. 
Kiss and Bedding 2003, Ita et al. 2004, Soszynski et al. 2005, 
Soszynski et al. 2007).

The most recent calibration of the various period-luminosity relations for
late-type giant variables has been presented by Soszynski et al. (2007).
Their relation for O-rich semi-regular variables of LMC takes the form
$K_s = -4.35(\log P - 2.0) + 11.25$, where $K_s$ band is that of the 2MASS
survey. Whitelock et al. (2008) have shown that any chemical abundance
effect on the $K$-band period-luminosity relation of Miras must be small.
Working with the revised Hipparcos parallaxes of van~Leeuwen~(2007),
Whitelock et al. (2008) have derived that period-luminosity relation of
O-rich Miras in our Galaxy has the same slope and it is 0.1 mag brighter
than the corresponding one for the LMC. We assume that a similar 0.1 mag
shift would make the Soszynski et al. (2007) relations applicable to the
O-rich SRa variables of our Galaxy. Adopting this 0.1 mag shift, a LMC
distance modulus of ($m$ - M)$_\circ$=18.39 (van Leeuwen et al. 2007), a LMC
reddening of $E_{B-V}$=0.06 (Mateo 1998), the extinction relation $A_{\rm
Ks} = 0.442 E_{B-V}$ for an M-type spectral distribution and a standard
$R_V$=3.1 extinction law (Fiorucci and Munari 2003), the distance to
IRC$-$10443 corresponding to the 85.5 ~day period is 0.5~kpc. Soszynski et
al. (2007) relation for the long secondary periods of O-rich red giants in
LMC takes the form $K_s= -4.41(\log P - 2.0) + 15.05$, and when applied
(with the same 0.1 mag shift as above) to the 620~day secondary periodicity
displayed by IRC$-$10443, it provides the same distance, 0.5~kpc, as
obtained for the 85.5 day main period. Such 0.5~kpc distance is adopted for
IRC$-$10443 in this paper.

\subsectionb{3.6}{On the nature of the long secondary period}
In spite of large investigation efforts, both observational and
theoretical, ``{\it the cause of the long secondary periods seen in cool
giants remains a mystery at the present time}'' as recently remarked by Wood
(2007).

It has been known for a long time that some semi-regular variables shows the
presence of a long secondary period (LSP) in their light-curves, typically
ten times longer than the primary pulsation period. This ratio for
IRC$-$10443 is 7.25. Lists of local giants displaying LSPs have
been published, among others, by Houk (1963), Mattei et al. (1997), Kiss et
al. (1999). Wood et al. (1999) found that $\sim$25\% of all variable AGB
star in LMC show LSPs. A similar fraction, $\simeq$30\%, of local semi-regular
variables has been found by Percy et al. (2004) to display LSPs.

\begin{wrapfigure}[29]{l}[0pt]{77mm}
\vbox{
\centerline{\psfig{figure=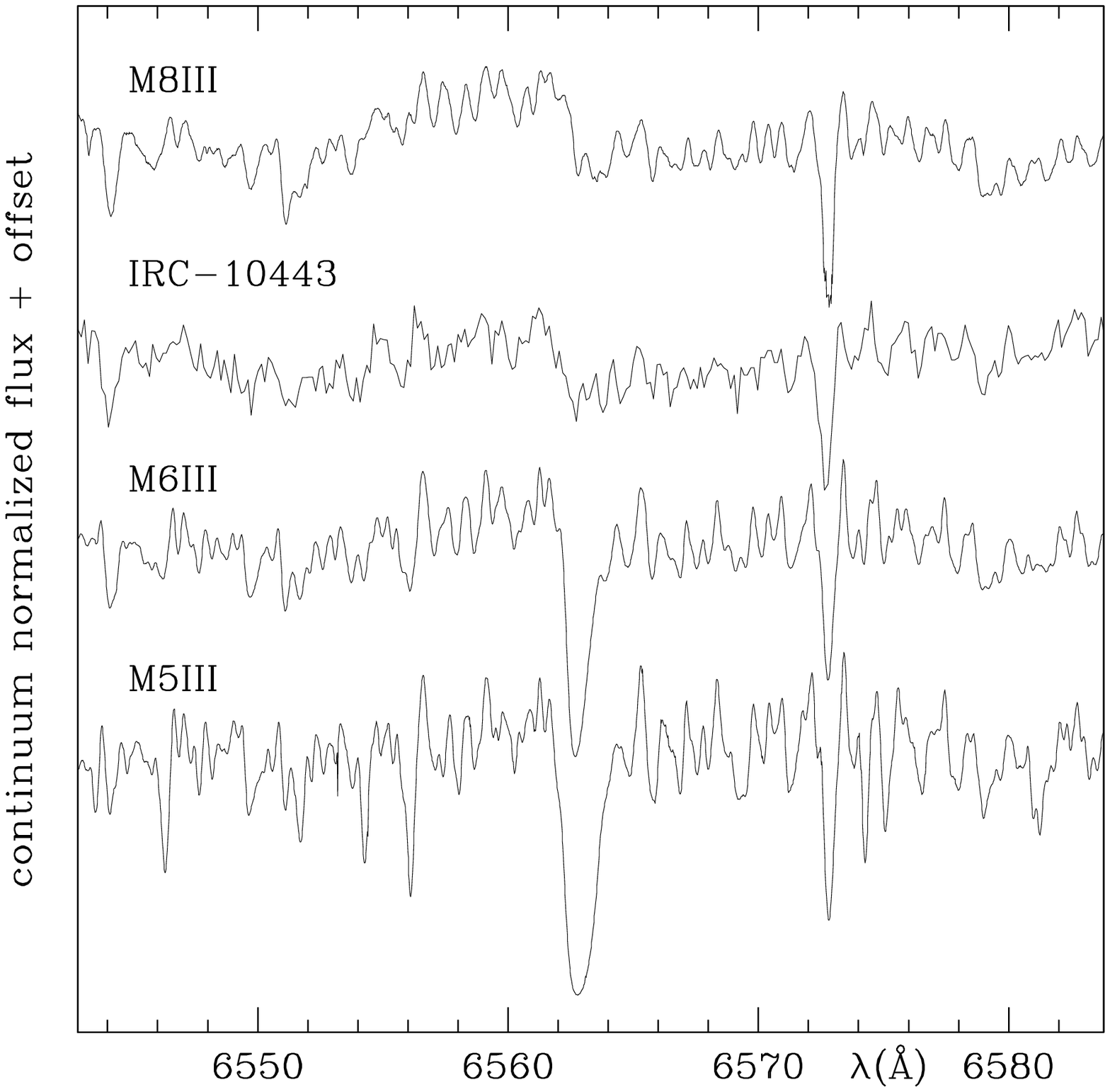,width=75truemm,angle=0,clip=}}
\captionc{5}{Portion, centered on H$\alpha$ and CaI 6572.8, of the high
      resolution 22.95 July 2008 Echelle spectrum of IRC$-$10443. Spectra of
      cool giants from Bagnulo et al. (2003) are plotted for reference. All 
      spectra are continuum normalized, offset in ordinates for better
      visibility, and shifted to 0.0 radial velocity.}
}
\end{wrapfigure}

The semi-regular variables appear to pulsate in the first overtone
(Lebzelter and Wood 2006), and thus it could be tempting to relate the LSPs
with pulsation in the fundamental mode. However, as found in theoretical
models by Fox and Wood (1982) and verified by observations (e.g. Kiss et al.
1999, Mattei et al. 1997), the ratio of fundamental to fist overtone periods
is close to two, ruling out that LSPs are due to pulsations in the
fundamental mode.  Established observational facts are that LSPs are
accompanied by radial velocity variations (of a lower amplitude than
observed for the primary period; Hinkle et al. 2002, Wood et al. 2004) and
by variation in intensity of the H$\alpha$ absorption (that could trace a
variable filling by an emission component of chromospheric origin; Wood et
al. 2004). In addition, cool variable giants showing LSPs rotate at similar
velocity and show similar dust-free IRAS colors as cool variable giants not
showing the LSPs (Olivier and Wood 2003). In RGB objects showing LSPs, the
period of associated radial velocity variations is twice the period of
photometric LSP variability (as expected in the case LSPs arise from
ellipsoidally distorted giants in binary systems; Adams et al. 2006), while
in AGB objects the period is the same (as expected in the case of
pulsations; Wood et al. 2004). Various explanations of the LSP phenomenon
have been proposed, but all have encountered some problems, as discussed by
Wood (2007, and references therein).

A striking feature displayed by IRC$-$10443 is illustrated in the bottom
panel of Figure~3, where the color variation is fitted with two sinusoids of
the same periods of those fitting the $V$ light-curve (cf Eq.~1 and the
top-panel of Figure~3). Their semi-amplitudes are 0.23~mag for the sinusoid
associated to the principal 85.5~day period, and 0.13~mag for the LSP, with
a mean $V-I_{\rm C}$=5.19. From Figure~3, it is evident that IRC$-$10443 is
bluest when it is at the maximum brightness along the LSP cycle, and reddest
when it is at the minimum brightness. This is the same pattern observed for
the principal 85.5~day period and strongly argues in favor of a pulsation
interpretation of LSP phenomenon, at least in IRC$-$10443. The simultaneous
presence of two pulsations also accounts for the dispersion of the points in
Figure~2 along the back-body curve. The dispersion would have been
significantly reduced if only one pulsation would have been present, as
confirmed by removing from observations one or the other of the sinusoidal
variation of the color and re-plotting Figure~2.

The variation in $V-I_{\rm C}$ color can be transformed into variation in
effective temperature of the underlying stellar photosphere using the
reference continuum energy distribution given by Fluks et al. (1994) along
the spectral sequence of M giants. The total amplitude of 0.46~mag observed
in $V-I_{\rm C}$ for the principal period, corresponds to a change in 0.73
spectral types around the M7III mean, and therefore to a variation from 3030
to 3175~K in effective temperature. The 0.26~mag total color amplitude of
the LSP would correspond to a change of 0.41 spectral types around the M7III
mean, meaning a variation from 3060 to 3145~K. If both pulsations are
supposed to occur at constant luminosity, the corresponding total excursion
in radius is $\sim$10\% for the principal 85.5~day period, and $\sim$6\% for
the LSP.

These excursions in radius are less than that inferred by Wood et al. (2004)
from radial velocity observations at optical wavelengths of a sample of
three hotter AGB stars characterized by longer LSP than IRC$-$10443. We
estimated the change in {\em effective} temperature and radius of the
underlying photosphere. If instead we had referred to a black-body fitting
of the observed optical spectrum of the star (re-shaped by the extremely
strong TiO molecular absorptions), the variation in {\em color} temperature
and radius of IRC$-$10443 would have been almost twice larger, 150~K and 
10~\%, respectively.

It is worth noticing that recent theoretical improvements in the treatment
of pulsation, like inclusion of time dependent turbulent convection (Olivier
and Wood 2006), are opening new modeling possibilities for pulsation
modes in cool giants. Important applications to the long lasting problem of what is
driving the LSPs could be obtained in the near future (cf. Wood 2006). To better
characterize the object and increase its interest as a test target for current
theories, we plan to continue a tight observational monitoring of IRC$-$10443 
over the next years, long enough to cover at least the whole next LSP period.

\vspace{5mm}
ACKNOWLEDGMENTS. We would like to thank P.A. Whitelock for useful comments
on the original version of the paper, the anonymous referee for helpful
suggestions, and S. Ciroi, F. Di Mille, S. Tomasoni, F. Moschini and M. Nave
for assistance during the observations.
  
\vskip10mm

\References

 \refb  Adams~E., Wood P.~R., Cioni~M.~R. 2006, MSAIt 77, 537
 \refb  Bagnulo~S., Jehin~E., Ledoux~C. et al. 2003, Messenger 114, 1
 \refb  Bessell~M.~S. 2000, PASP 112, 961
 \refb  Brand~J., Blitz~L. 1993, A\&A 275, 67
 \refb  Fiorucci~M., Munari~U. 2003, A\&A 401, 781
 \refb  Fitzgerald~M.~P. 1970, A\&A 4, 234
 \refb  Fluks~M.~A., Plez~B., The~P.~S. et al. 1994, A\&AS 105, 311
 \refb  Fox~M.~W., Wood~P.~R. 1982, ApJ 259, 198
 \refb  Ita~Y., Tanabe~T., Matsunaga~N. et al. 2004, MNRAS 353, 705
 \refb  Johnson~H.~L. 1966, ARA\&A 4, 193
 \refb  Hansen~O.~L., Blanco~V.~M. 1975, AJ 80, 1011
 \refb  Hinkle~K.~H., Lebzelter~T., Joyce~R.~R., Fekel~F.~C. 2002, AJ 123, 1002
 \refb  Houk~N. 1963, AJ 68, 253
 \refb  Kiss~L.~L., Szatmáry~K., Cadmus~R.~R., Mattei~J.~A.  1999, A\&A 346, 542 
 \refb  Kiss~L.~L., Bedding,~T.~R. 2003, MNRAS 343, L79
 \refb  Kwok~S., Volk~K., Bidelman~W.~P. 1997, ApJS 112, 557
 \refb  Landolt~A.~U. 1992, AJ 104, 340
 \refb  Lebzelter~T., Wood~P.~R. 2006, MSAIt 77, 55
 \refb  Lee~T.~A. 1970, ApJ 162, 217
 \refb  Madsen~G.~J., Reynolds~R.~J. 2005, ApJ 630, 925
 \refb  Mateo~M.~L. 1998, ARA\&A 36, 435
 \refb  Mattei~J.~A., Foster~G., Hurwitz~L.~A. et al. 1997, in Hipparcos - Venice '97, ESA SP-402, 269
 \refb  Mikulasek~Z., Graf~T. 2005, CoSka 35, 83
 \refb  Nassau~J.~J., Blanco~V.~M., Cameron~D.~M. 1956, ApJ 124, 522
 \refb  Neckel~H. 1958, ApJ 128, 510
 \refb  Neugebauer~G., Leighton~R.~B. 1969, Two-Micron Sky Survey. A Preliminary Catalogue, NASA SP, Washington
 \refb  Olivier~E.~A, Wood~P.~R. 2003, ApJ 584, 1035
 \refb  Olivier~E.~A, Wood~P.~R. 2006, MSAIt 77, 515
 \refb  Panchuk~V.~E. 1978, SvAL 4, 201 
 \refb  Percy~J.~R., Bakos~A.~G., Besla~G. et al. 2004, in Variable Stars in the Local Group, IAU Colloquium 193, D.W. Kurtz and K.R. Pollard eds., ASPC 310, 348
 \refb  Price~S.~D., Murdock~T.~L. 1983, The Revised AFGL I.R. Sky Survey, Catalog and Supplement, Air Force Geophysics Lab. AFGL-IR-83-0161 
 \refb  Reichen~M., Lanz~T., Golay~M., Huguenin~D. 1990, Ap\&SS 163, 275
 \refb  Soszynski~I., Udalski~A., Kubiak~M. et al.  2005, AcA 55, 331
 \refb  Soszynski~I., Dziembowski~W.~A., Udalski~A. et al.  2007, AcA 57, 201
 \refb  van~Leeuwen~F. 2007, Hipparcos: The New Reduction of the Raw Data, Springer-Verlag
 \refb  van~Leeuwen~F., Feast~M.~W., Whitelock~P.~A., Laney~C.~D. 2007, MNRAS 379, 723
 \refb  Yamashita~Y., Nariai~K., Norimoto~Y. 1977, An Atlas of Representative Stellar Spectra, Univ. of Tokio Press
 \refb  Whitelock~P.~A. 1996, in Light Curves of Variable Stars, C.Sterken and C.Jaschek eds., Cambridge Univ. Press
 \refb  Whitelock~P.~A., Feast~M.~W., van Leeuwen~F. 2008, MNRAS 386, 313
 \refb  Wood~P.~R., Sebo~K.~M. 1996, MNRAS 282, 958
 \refb  Wood~P.~R., Alcock~C., Allsman~R.~A. et al. 1999, in
              Asymptotic Giant Branch Stars, T.~Le~Bertre, A.~Lebre and 
              C.~Waelkens eds., IAU Symp 191, 151
 \refb  Wood~P.~R., Olivier~E.~A., Kawaler~S.~D. 2004, ApJ 604, 800
 \refb  Wood~P.~R. 2006, MSAIt 77, 76
 \refb  Wood~P.~R. 2007, in The 7th Pacific Rim Conference on Stellar Astrophysics, Y.~W.~Kang et al. eds., ASPC 362, 234
 \refb  Zacharias~N., Monet~D.~G., Levine~S.~E. et al. 2004, AAS 205, 4815

\end{document}